\documentclass[aps,prd,twocolumn,superscriptaddress,amsfont,graphicx,nofootinbib,preprintnumbers]{revtex4}%
\usepackage{color,graphicx,epsfig}
\usepackage{ifpdf}
\usepackage{amsmath}
\usepackage{bm}
\usepackage{color}
\usepackage[english]{babel}
\usepackage{graphicx}%
\usepackage{amsfonts}%
\usepackage{amssymb}
\usepackage{braket}
\usepackage{hyperref}
\usepackage{subfigure}
\newenvironment{Eqnarray}{\arraycolsep 0.14em\begin{eqnarray}}{\end{eqnarray}}
\def\beqa{\begin{Eqnarray}}
\def\eeqa{\end{Eqnarray}}

\definecolor{nicered}{rgb}{0.7,0.1,0.1}
\definecolor{nicegreen}{rgb}{0.1,0.5,0.1}
\hypersetup{colorlinks,citecolor= nicegreen,linkcolor= nicered}

\newcommand{\beq}{\begin{equation}}
\newcommand{\eeq}{\end{equation}}
\newcommand{\bea}{\begin{eqnarray}}
\newcommand{\eea}{\end{eqnarray}}

\newcommand{\ahm}{\alpha_{vh}}

\begin{document}

\def\LjubljanaFMF{Faculty of Mathematics and Physics, University of Ljubljana,
 Jadranska 19, 1000 Ljubljana, Slovenia }
\def\LjubljanaIJS{Jo\v zef Stefan Institute, Jamova 39, 1000 Ljubljana, Slovenia}
\def\CERN{CERN, Theory Division, CH-1211 Geneva 23, Switzerland}
\def\Weizmann{Department of Particle Physics and Astrophysics,
Weizmann Institute of Science, Rehovot, Israel 761000}

\title{The phenomenology of the di-photon excess and $h\to\tau\mu$ within 2HDM}

\author{Aielet Efrati}
\email[Electronic address:]{aielet.efrati@weizmann.ac.il}
\affiliation{\Weizmann}

\author{Jernej F.\ Kamenik}
\email[Electronic address:]{jernej.kamenik@cern.ch}
\affiliation{\CERN}
\affiliation{\LjubljanaIJS}
\affiliation{\LjubljanaFMF}

\author{Yosef Nir}
\email[Electronic address:]{yosef.nir@weizmann.ac.il}
\affiliation{\Weizmann}


\begin{abstract}
The diphoton excess around $m_S=750$ GeV observed at ATLAS and CMS can be interpreted as coming from $S=H$ and $A$, the neutral components of a second Higgs doublet. If so, then the consistency of the light Higgs decays with the Standard Model predictions provides upper bounds on the rates of $S\to VV,\ hZ,\ hh$ decays. On the other hand, if $h\to\tau\mu$ decay is established, then a lower bound on the rate of $S\to\tau\mu$ decay arises. {Requiring that $\Gamma_S\lesssim45$ GeV gives both an upper and a lower bound on the rotation angle from the Higgs basis $(\Phi_v,\Phi_A)$ to the mass basis $(\Phi_h,\Phi_H)$}.  The charged scalar, with $m_{H^\pm}\simeq750$~GeV, is produced in association with a top quark, and can decay to $\mu^\pm\nu$, $\tau^\pm\nu$, $tb$ and $W^\pm h$.
\end{abstract}

\maketitle

\section{Introduction}\label{sec:intro}
Two of the most interesting measurements at ATLAS/CMS are the search for high mass diphoton resonances, and the search for the lepton flavor violating (LFV) decay $h\to\tau\mu$. The searches for diphoton resonances find an excess around $m_{\gamma\gamma}=750$~GeV \cite{Aaboud:2016tru,Khachatryan:2016hje}, with (c.f.~\cite{Franceschini:2015kwy,Franceschini:2016gxv} and references within)
\beq\label{eq:StoAA}
\sigma_{\gamma\gamma}^{13} \equiv \sigma(pp\to S)\times {\rm BR}(S\to\gamma\gamma)\approx(8\pm2)\,{\rm fb}\,,
\eeq
at  $\sqrt{s}=13\ {\rm TeV}$\,.
The searches for $h\to\tau\mu$ yield the following ranges \cite{Khachatryan:2015kon,Aad:2016blu}:
\begin{equation}\label{eq:BRmutauc}
{\rm BR}(h\to\tau\mu)=\left\{\begin{matrix}(8.4^{+3.9}_{-3.7})\times10^{-3}&{\rm CMS,}\\
(5.3\pm5.1)\times10^{-3}&{\rm ATLAS.} \end{matrix}\right.
\end{equation}

A diphoton resonance $S$ can be related to a new state of spin-2 or spin-0. Focussing on the scalar option, it could be an $SU(2)$-singlet or doublet. The latter possibility is suggestive that $S$ can be the result of a pair of neutral scalars, $H$ and $A$. If $H$ and $A$ are close in mass but not degenerate, that can explain the ATLAS result of large width.

A branching ratio of $h\to\tau\mu$ of order a percent implies that the decay rate is not much smaller than the $h\to\tau\tau$ decay rate. The $h\to\tau\mu$ decay is, however, a flavor changing neutral current process which, within the Standard Model (SM) and many of its extensions (such as the minimal supersymmetric SM), is loop suppressed. Thus, ${\rm BR}(h\to\tau\mu)\not\ll{\rm BR}(h\to\tau\tau)$ is suggestive that $h\to\tau\mu$ proceeds at tree level. The two most plausible relevant extensions of the SM are vector-like leptons and multi Higgs doublets. The former framework leads, generally, to unacceptably large $Z\to\tau\mu$ and $\tau \to \mu \gamma$ decay rates~\cite{Falkowski:2013jya}, leaving the two Higgs doublet model as the simplest model that can account for ${\rm BR}(h\to\tau\mu)={\cal O}(0.01)$.

In this work, we entertain the possibility that the experiments will establish both ${\rm BR}(h\to\tau\mu)={\cal O}(0.01)$ and a scalar resonance $S$ with $m_S\approx750$ GeV and that, furthermore, $h$ and $S$ are the neutral scalars of a two Higgs doublet model (2HDM). If this is the case, then measurements of the $h$ couplings provide testable predictions for the $S$ couplings.

It is convenient for our purposes to define the rotation angle $\ahm$ from the Higgs basis, where one doublet carries the electroweak breaking vacuum expectation value (VEV) and the other is VEV-less, to the mass basis of the CP-even scalars, where one doublet contains $h$ and the other $H$. The angle $\ahm$ is related to the conventional $\alpha-\beta$ with $\ahm=\pi/2+\alpha-\beta$.

The upper bounds on deviations of the $hVV$ couplings from the $h_{\rm SM}VV$ couplings imply that $\tan\ahm\lesssim1$.
We further define
\begin{align}\label{def:rsxy}
R^{S/h}_{XY}&\equiv \frac{\Gamma(S\to XY)}{\Gamma(h\to XY)}\ \ \ (S=H,A),\\
R^{\phi_i/h_{\rm SM}}_{XY}&\equiv \left.\frac{\Gamma(\phi_i\to XY)}{\Gamma(h_{\rm SM}\to XY)}\right|_{m_{h_{\rm SM}}=m_{\phi_i}}\ \ \ (\phi_{i}=H,A,h).\nonumber
\end{align}

We distinguish four classes of $S=H,A$ decay modes, according to their dependence on the 2HDM parameters (beyond their dependence on $m_S$):
\begin{enumerate}
\item $S\to\tau^\pm\mu^\mp$: $R^{S/h}_{\tau\mu}$ depends on only $\ahm$. An upper bound on $\sin\ahm$ gives a lower bound on $R^{S/h}_{\tau\mu}$.
\item $H\to VV$ ($V=W,Z$): $R^{H/h}_{VV}$ depends on only $\ahm$. An upper bound on $\sin\ahm$ gives an upper bound on $R^{S/h}_{VV}$.
\item $S\to f\bar f$: $R^{S/h}_{f\bar f}$ depends on $\ahm$ and an additional, flavor-dependent parameter $\eta_f$, to be defined below.
\item $S\to\gamma\gamma$ and $S\to gg$: $R^{S/h}_{\gamma\gamma.gg}$ depend on the UV completions of the 2HDM (as will be discussed in the following).
\end{enumerate}
This classification makes the special significance of $h\to\tau\mu$ manifest: The combination of measuring $\Gamma(h\to\tau\mu)\neq0$ and an upper bound on the deviation of $\Gamma(h\to VV)$ from its SM value provide a lower bound on $\Gamma(S\to\tau\mu)$.

The diphoton excess requires that there are new physics contributions to the $S\gamma\gamma$ coupling (c.f.~\cite{Angelescu:2015uiz}).
The signals depend, however, not only on the decay rates but also on the cross section for production.
In this context, it is natural to assume that the second doublet couples most strongly to the third generation quarks. Coupling to the top quark will lead to gluon-gluon fusion (ggF) production of $S$. Coupling to the bottom quark will generate $S$ via $b\bar b$ production. In either case, strong statements about the event rates for various final states of $S$ decays can be made. The purpose of this work is to obtain these predictions, and to study in detail the phenomenology of the charged scalar in this framework, assuming $m_{H^+}\approx750$~GeV.

The idea to interpret both $S(750)\to\gamma\gamma$ and $h\to\tau\mu$ in the framework of a 2HDM was first made in Ref.~\cite{Bizot:2015qqo}. (See also~\cite{Han:2016bvl,Bolanos:2016aik}.) In this paper we extend their work in the following ways:
\begin{itemize}
\item We analyse the production and decays of the charged Higgs $H^\pm$ in this framework.
\item We allow generic couplings of the two Higgs doublets to $t\bar t$ or $b\bar b$.
\item We stay agnostic to the mechanism that generates the di-photon coupling of $S$ and are consequently led to a different evaluation of the $\tau\to\mu\gamma$ constraints.
\item We include the $A\to hZ$ and $H\to hh$ decays in our analysis.
\end{itemize}

The plan of this paper is as follows. We introduce our theoretical framework in Section \ref{sec:model}. The explicit expressions for the $S$ decay rates are presented in Section \ref{sec:rates}. We review the experimental constraints that apply to this model in Section \ref{sec:constraints}. The numerical results are presented in Section \ref{sec:results}. In Section \ref{sec:charged} we describe the phenomenology of the charged Higgs in our framework. We conclude in Section \ref{sec:conclusions}.

\section{The model}\label{sec:model}
We work in a general 2HDM. One is free to rotate the two Higgs doublets ($\Phi_{1,2}$) into a basis where only one obtains a VEV:
\beq
\Phi_v = \left ( \begin{array}{c} G^+ \\ \frac{1}{\sqrt 2} (v+h_1 + i G^0) \end{array}
 \right ) \,, \quad \Phi_A = \left ( \begin{array}{c} H^+ \\ \frac{1}{\sqrt 2} (h_2 + i h_3) \end{array}
 \right )\,.
 \eeq

In the CP conserving limit (assuming all parameters in the scalar potential to be real) the CP-odd pseudoscalar $h_3$ does not mix with the other neutral states and thus forms a mass eigenstate $h_3 \equiv A^0$. The two CP even scalars $h_{1,2}$, on the other hand, do mix to form the mass eigenstates $h$ and $H^0$:
\beq
\left ( \begin{array}{c} h \\ H^0 \end{array} \right ) =  \left ( \begin{array}{cc}\cos\ahm & \sin\ahm \\ -\sin\ahm & \cos\ahm \end{array} \right )  \left ( \begin{array}{c} h_1 \\ h_2 \end{array} \right )\,.
\eeq
We identify $h$ with the observed Higgs boson at $m_h\simeq125$~GeV. We further associate $H^0$ and $A^0$ with the di-photon excess at the LHC at $m_S\approx750$~GeV, in which case
\beq
m_{A^0} \simeq m_{H^0} \simeq m_{H^+} \equiv m_S\,,
\eeq
up to possible corrections of order $v^2/m_S^2 \simeq 0.1$\,. The Yukawa couplings of the neutral mass eigenstate scalars to SM fermions are given by
\begin{align}
\mathcal L_{f} &= - \bar f_L^i f_R^j \left[ h\left( \frac{m_i}{v} \delta_{ij} \cos\ahm + \frac{\eta_{ij}^{f}}{\sqrt 2} \sin\ahm \right) \right.  \nonumber\\
 & \left.+ H^0\left( -\frac{m_i}{v} \delta_{ij} \sin\ahm + \frac{\eta_{ij}^{f}}{\sqrt 2} \cos\ahm \right) + i A^0 \frac{\eta_{ij}^{f}}{\sqrt 2}  \right] \,.
 \end{align}
The $\eta^f$ matrices ($f=u,d,e$) are the Yukawa matrices of $\Phi_A$ in the $f$ mass basis. For flavor diagonal Yukawa couplings, it is convenient to define
\beq
\hat\eta_x\equiv\frac{\eta_{xx} v}{\sqrt2 m_x}\,.
\eeq
As concerns the couplings of $S$ to di-photons, it was proven
(c.f.~\cite{Angelescu:2015uiz, Aloni:2015mxa, Gupta:2015zzs}) that there must be contributions from new degrees of freedom beyond the 2HDM. We parameterize these contributions by writing the following effective couplings (still assuming CP conservation):
\beq
\mathcal L_\gamma=\frac{\alpha_{\rm EM}}{\pi v} c_\gamma h_2 F^{\mu\nu}F_{\mu\nu}+\frac{\alpha_{\rm EM}}{\pi v} \tilde c_\gamma A^0 F^{\mu\nu}\tilde F_{\mu\nu}\,,
\eeq
with $\tilde F_{\mu\nu}=\epsilon_{\mu\nu\alpha\beta}F^{\alpha\beta}/2$. The low energy theorems and the ABJ anomaly imply, in the CP conserving case, that $c_\gamma=\frac23\tilde c_\gamma$. Given the consistency of the $h\to\gamma\gamma$ decay rate with the SM prediction we assume that the beyond-2HDM contributions to $c_\gamma^h$ are negligible. This situation is easily realized, for instance, with additional heavy vector-like leptons. In this case
\beq
c_\gamma^A\simeq\frac32 c_\gamma^H\,,
\eeq
with corrections of order $\sin\ahm^2$.

In principle, the UV degrees of freedom might generate additional operators at the high scale. While the exact determination of the full EFT depends on the details of the high scale theory, some general conclusions can be drawn under mild assumptions. To do so, we assume that at the high scale $\Lambda$, only the operators
\beq
\mathcal L \ni \sum_{ij=1,2} (c_{ijBB} Q_{ijBB}   + c_{ij\tilde BB}Q_{ijB\tilde{B}}) + \rm h.c.\,,
\eeq
\begin{align}
Q_{ijBB}&\equiv H^\dagger_iH_jB_{\mu\nu}B^{\mu\nu} , \quad  Q_{ijB\tilde B}\equiv H^\dagger_iH_jB_{\mu\nu}\tilde B^{\mu\nu}\,,
\end{align}
are generated with real coefficients $c_{ijBB}, c_{ij\tilde BB}$. The corresponding operators involving $SU(2)_L$ gauge bosons are absent, for instance, if the new degrees of freedom are $SU(2)_L$ singlets.

As concerns the UV completion of the this EFT, it was pointed out~\cite{Goertz:2015nkp,Bertuzzo:2016fmv} that a large multiplicity of vector-like quarks leads to instabilities in the scalar potential at relatively low scales. Staying agnostic to the exact details of the UV dynamics, we admit the need for non-generic high scale spectrum or couplings to avoid such instabilities, while its exact determination is beyond the scope of this work. Perturbativity of the gauge couplings, however, is ensured at the TeV scale in this scenario\cite{Franceschini:2015kwy}.

A priori, our setup introduces large number of parameters: the rotation angle $\sin\ahm$, the $A$ coupling to fermions $\eta_{ij}$, and the effective coupling to photons, $c_\gamma$, generated by the high scale dynamics. As for the couplings to quarks, we take an ansatz in which only the couplings to third generation quarks are significant. For simplicity we consider two scenarios, $\hat\eta_t\neq0$ or $\hat\eta_b\neq0$, where the extension in which both exist is straightforward.

The LFV couplings $\eta_{\mu\tau,\tau\mu}$ are determined using the $h\to\tau\mu$ excess, Eq.~\eqref{eq:BRmutauc} (and the corresponding Eq.~\eqref{eq:htaumu}). The effective coupling to photons is determined using the di-photon excess, Eq.~\eqref{eq:StoAA}. (We use the median value $\sigma^{13}_{\gamma\gamma}=5{\rm ~fb}$ from a global fit to CMS and ATLAS 8~TeV and 13~TeV data in the case of $gg$ or $b\bar b$ mediated production of a wide di-photon resonance~\cite{Kamenik:2016tuv}.) We are then left with two free parameters for each scenario we consider. These are:
\begin{align}
{\rm Case ~I:~}&\sin\ahm\,,\hat\eta_t\,,\nonumber\\
{\rm Case ~II:~}&\sin\ahm\,,\hat\eta_b\,.
\end{align}

\section{Scalar decay rates}
\label{sec:rates}
We now write explicit expressions for decay rates corresponding to the classification given in the Sec.~\ref{sec:intro}.

1. The $h\to\tau\mu$ decay rate is given by
\beq
\Gamma(h\to \tau^\pm \mu^\mp)=\frac{m_h}{16\pi} \sin^2\ahm \left(  |\eta_{\tau\mu}|^2 +  |\eta_{\mu\tau}|^2 \right)\,,
\eeq
The recent CMS indications of ${\rm BR}(h \to \tau\mu)\equiv{\rm BR}(h \to \tau^+\mu^-)+{\rm BR}(h \to \tau^-\mu^+)=(0.84^{+0.39}_{-0.37})$ then imply
\begin{align}\label{eq:htaumu}
|\sin\ahm| \sqrt{|\eta_{\tau\mu}|^2+|\eta_{\mu\tau}|^2} \simeq 0.0037\,.
\end{align}
We note that this result hold, to a good approximation, only for $|\sin\ahm\hat\eta_t|\ll1$, in which case the production cross-section of $h$ is very close to the SM cross-section (c.f.~\cite{Dorsner:2015mja}). (Our numerical analysis includes the full corrections to this experimental interpretation.)
The $H,A\to\tau\mu$ decay rates read
\beq
R^{H/h}_{\tau\mu}= \frac{m_{H}}{m_h}{\cot}^{2}\ahm\,, \ \ \
R^{A/h}_{\tau\mu}= \frac{m_{A}}{m_h}{\csc}^{2}\ahm\,.
\eeq

2. The relevant decay rates involving the electroweak vector bosons (and, equivalently, the vector boson fusion production rates) read
\begin{align}
R^{h/h_{\rm SM}}_{VV}&= \cos^2\ahm\,,\\
R^{H/h_{\rm SM}}_{VV}&= \sin^2\ahm\,, \nonumber\\
R^{A/h_{\rm SM}}_{VV}&= 0\,. \nonumber
\end{align}
The relevant decay modes involving the light Higgs read\footnote{We quote here the leading order result for the $H\to hh$ decay rate. This result might change significantly if $Z_3$, as defined in Ref.~\cite{Bernon:2015qea}, is $\mathcal{O}(30)$, which requires, in turn, fine tuning in the mass term of $h_2$. We therefore ignore this possibility in the following.}
\begin{align}
\Gamma(A\to Zh)&=\frac{G_F m_A^3 \sin^2\ahm}{8\sqrt{2}\pi}\lambda\left[m_h^2,m_Z^2,m_A^2\right]^{3/2}\,, \\
\Gamma(H\to hh)&=\frac{9G_F m_H^3 \sin^2\ahm}{16\sqrt{2}\pi}\beta[2m_h,m_H]\,,
\end{align}
with $\lambda[x,y,z]=\left(1-x/z-y/z\right)^2-4xy/z^2$ and $\beta[x,y]=\sqrt{1-x^2/y^2}$.

3. For flavor-diagonal decays into the SM fermions, we have
\begin{align}
R^{h/h_{\rm SM}}_{f\bar f}&= \left|\cos\ahm+\sin\ahm\hat\eta_f\right|^2\,,\\
R^{H/h_{\rm SM}}_{f\bar f}&= \left|-\sin\ahm+\cos\ahm\hat\eta_f\right|^2\,, \\
R^{A/h_{\rm SM}}_{f\bar f}&= \left|\hat\eta_f\right|^2\,.
\end{align}

4. For the decays into two gluons, dominated by the top loop (and, equivalently, for the gluon-gluon fusion production rates):
\begin{align}
R^{h/h_{\rm SM}}_{gg}&= \left|\cos\ahm+\sin\ahm\hat\eta_t\right|^2\,,\\
R^{H/h_{\rm SM}}_{gg}&= \left|-\sin\ahm+\cos\ahm\hat\eta_t\right|^2\,, \\
R^{A/h_{\rm SM}}_{gg}&= r^{A/H}_{QCD} \left|[\mathcal P(\tau_{t/A})/\mathcal S(\tau_{t/A})]\hat\eta_t\right|^2\,,
\end{align}
where $r^{A/H}_{QCD} \simeq 0.88$ takes into account the somewhat different higher order QCD effects for scalar and pseudo-scalar fields~\cite{Harlander:2005rq}, and where $\tau_{t/A}=4m_t^2/m_A^2$ and
\beq
\mathcal P(\tau) = \arctan^2(1/\sqrt{\tau-1})\,,\quad \mathcal S(\tau) = 1+ (1-\tau)\mathcal P(\tau)\,.
\eeq
For the decays into two photons, we have
\begin{align}
\Gamma(S\to\gamma\gamma)&\simeq\frac{\alpha_{\rm EM}^2}{4\pi^3}\frac{m_H^3}{v^2}|C^S_\gamma|^2\,,
\end{align}
with
\begin{align}
C^H_{\gamma}&=\cos \alpha_{vh} c_\gamma+\left(-\sin\ahm+\cos\ahm\hat\eta_t\right)c_t-\sin\ahm c_W\,,\nonumber\\
C^A_{\gamma}&=\frac{3}{2}c_\gamma+\hat\eta_t \tilde c_t\,,
\end{align}
where $c_t\simeq0.1+0.2i$, $\tilde c_t\simeq0.04+0.3i$ and $c_W\simeq-0.2-0.2i$ are the LO contributions from the top quark and $W$ boson loops. The QCD corrections modify $c_t, \tilde c_t$ by $\mathcal O(10\%)$~\cite{Harlander:2005rq}. Numerically, these effects are always negligible relative to the $c_\gamma$ contribution.

In principle $h\to gg$ and $h\to\gamma\gamma$ transitions can obtain contributions also from other SM fermion loops if the corresponding $\hat\eta_{f}$ are sizeable.
In practice, light SM fermion contributions to $H^0, A^0 \to gg, \gamma\gamma$ decays are completely negligible as they are suppressed by small loop functions. The $b$ quark loop is rendered negligible in the region of $\hat\eta_b$ which is allowed by the Higgs data. The charged Higgs contribution to the di-photon rate is small as it is suppressed by $\sim\ahm v^2/m_S^2$.

The EFT operators we consider contribute also to $S\to Z\gamma$ and $S\to ZZ$ decays. These contributions, however, are suppressed by $\tan^2\theta_W/2$ and $\tan^4\theta_W$ respectively, relative to the the di-photon ones. The current sensitivities of the direct searches in these channels are insufficient to probe the EFT contributions to these decays. We therefore neglect these in the following.


\section{Experimental constraints}
\label{sec:constraints}

Here we detail the various experimental constraints we consider. We incorporate those, for the two cases specified above, in the next section, which presents our numerical results.

\subsection{Direct  $S$ searches}

Since the only $S$ decay to have been observed is $S\to\gamma\gamma$, the various direct searches for $S$ constrain its various decay widths. At 750~GeV they read
\begin{table}[h!]
\centering
\label{tab:direct}
\begin{tabular}{|c|l|l|}
\hline
Bound    						& Ref.			   &  Definition						               \\ \hline \hline
$\sigma_{\tau\mu}^8<20{\rm \,fb}$	& \cite{Aad:2015pfa}    &  $\sigma_8(pp\to S)\times{\rm BR}(S\to \tau\mu)$	       \\
$\sigma_{ZZ}^8<12{\rm \,fb}$		& \cite{Aad:2015kna}    &  $\sigma_8(pp\to H)\times{\rm BR}(H\to ZZ)$	       \\
$\sigma_{Z\gamma}^{8+13}<12{\rm \,fb}$ & \cite{CMS:2016ion} & $\left[0.12\sigma_{13}(pp\to S)+0.88\sigma_8(pp\to S)\right]$ \\
& & $\times{\rm BR}(S\to Z\gamma)$ \\
$\sigma_{t\bar t}^8<0.6{\rm \,pb}$ & \cite{Chatrchyan:2013lca} & $\sigma_8(pp\to S)\times{\rm BR}(S\to t\bar t)$ \\
$\sigma_{Zh}^8<19 {\rm \,fb}$ & \cite{AZh} & $\sigma_{8}(pp\to A)\times{\rm BR}(A\to Zh)$ \\
& & $\times{\rm BR}(h\to b\bar b)$ \\
$\sigma_{hh\rightarrow4b}^{13}<52 {\rm \,fb}$ &\cite{Hhh}& $\sigma_{13}(pp\to H)\times{\rm BR}(H\to hh)$\\
& & $\times{\rm BR}(h\to b\bar b)^2$ \\
\hline \hline
\end{tabular}
\end{table}

In addition, we consider ATLAS best fit value for the total width, $\Gamma_S/m=6\%$, as an upper bound for the total widths of $H$ and $A$.

If the production cross section of $S$ is small, a large $c_\gamma$ is required to accommodate the correct di-photon signal. In such a case, the photon fusion production becomes significant, resulting in some tension with the 8~TeV di-photon bounds. In particular, small ratios of signal strengths $r^{13/8}_{\gamma\gamma} \equiv \sigma_{\gamma\gamma}^{13} / \sigma_{\gamma\gamma}^{8}$ are disfavored at $2\sigma$ ($3\sigma$) for $r^{13/8}_{\gamma\gamma}<3(2)$~\cite{Kamenik:2016tuv}. We use MSTW2008NLO PDF set~\cite{Martin:2010db} to estimate the photon fusion contribution to the $S$ production which gives $r^{13/8}_{\gamma\gamma}=1.9$ for pure photon fusion~\cite{Franceschini:2015kwy}. 

We further verify that partial wave unitarity of a di-photon scattering process is not violated in the allowed parameter space, as analyzed in~\cite{DiLuzio:2016sur}.

\subsection{EW precision tests}

The 2HDM scenario modifies the vacuum polarization of the EW gauge bosons, changing the values of the oblique parameters~\cite{Peskin:1991sw} from their SM predictions. In the limit $m_{H^+}=m_A=m_H$, these corrections are proportional to $\sin^2\ahm$. Using the results of~\cite{Baak:2011ze} (and references within), we find
\begin{align}
S=0.10\sin^2\ahm\,,\;\;T=-0.20\sin^2\ahm\,,\;\;U=-0.03\sin^2\ahm\,,
\end{align}
to be confronted with the Gfitter results~\cite{Baak:2014ora},
\begin{align}
S=0.05\pm0.11\,,\;\;\;T=0.09\pm0.13\,,\;\;\;U=0.01\pm0.11\,.
\end{align}
For $\sin\ahm\lesssim0.3$, these are satisfied within $95\%$ C.L..

Possible RGE mixing between $Q_{11BB}$ and $Q_{11WB}$ induces an additional correction to the $S$ parameter. If, for instance, at 1 TeV $c_{12BB}\sim c_{11BB}$ we find that $S\sim0.1$ is generated at the weak scale for the relevant $c_{12BB}$ needed for the di-photon signal. It is plausible, however, to realize a UV model in which $c_{11BB}=0$ or that cancelations with other operators arise at the weak scale. Staying agnostic to the complete realization at the UV, and since $S\sim0.1$ is still compatible with LEP results within the $1\sigma$ level, we do not consider this contribution to the $S$ parameter.

\subsection{$\tau\to\mu\gamma$}

The presence of $\eta_{\tau\mu,\mu\tau}$ leads, in general, to constraints coming from the $\tau\to\mu\gamma$ decay searches, currently implying ${\rm BR}(\tau\to\mu\gamma) < 4.4 \times 10^{-8} \,@\, 90\%$ C.L.~\cite{Aubert:2009ag}. The radiative decays receive important contributions at both the one- and two-loop levels. While the two loop contributions depend on $\eta_{\mu\tau,\tau\mu}$ and $c_\gamma$ which, within our framework, are determined by experiment, the one-loop contributions depend also on $\eta_{\tau\tau}$ and $\eta_{\mu\mu}$. To understand the possible impact of the constraints from $\tau\to\mu\gamma$ on $S\to\tau\mu$, we consider here the simple case where $\eta_{\tau\tau}=\eta_{\mu\mu}=0$. This case is representative of the bulk of parameter space, and does not introduce accidental cancelations. Given, however, that cancelations among the various scalar mediated contributions might occur, we present in the next section our results for the case that there are no significant constraints arising from $\tau\to\mu\gamma$.

Using the relevant effective Lagrangian
\beq
\mathcal L_{\tau\to \mu\gamma} = \frac{em_\tau}{8\pi^2}\left[c_L \bar \mu (\sigma \cdot F) P_L \tau +  c_R \bar \mu (\sigma \cdot F) P_R \tau + \rm h.c.\right]\,,
\eeq
one obtains
\beq
\Gamma(\tau \to \mu \gamma) = \frac{\alpha_{\rm EM} m_\tau^5}{64\pi^4} \left( |c_L|^2 + |c_R|^2 \right)\,,
\eeq
where $c_{L,R}$ is evaluated at the scale $m_\tau$. In the following we neglect the weak running of the dipole operator.

The one- and two-loop contributions to $c_{L,R}$ in the framework of 2HDM have been studied extensively in the literature~\cite{Blankenburg:2012ex, Harnik:2012pb,Crivellin:2013wna, Omura:2015nja}. At the one-loop order, they read
\begin{align}
c_{L,R}^{1-{\rm loop},h} &= \frac{m_\tau\eta_{\mu\tau,\tau\mu}}{\sqrt2 vm_h^2}  s_{\ahm} \left(c_{\ahm} + \hat\eta_\tau s_{\ahm}\right) \nonumber\\
&\times \left( -\frac{1}{3} + \frac{1}{4} \log \frac{m_h^2}{m_\tau^2} \right)\,, \\
c_{L,R}^{1-{\rm loop},H} &= \frac{m_\tau\eta_{\mu\tau,\tau\mu}}{\sqrt2 vm_H^2}  c_{\ahm} \left(-s_{\ahm} + \hat\eta_\tau c_{\ahm}\right) \nonumber\\
&\times \left( -\frac{1}{3} + \frac{1}{4} \log \frac{m_H^2}{m_\tau^2} \right)\,, \\
c_{L,R}^{1-{\rm loop},A} &= -\frac{m_\tau\eta_{\mu\tau,\tau\mu}}{2\sqrt2 vm_A^2}   \hat\eta_\tau \times \left(-\frac{5}{12}+\frac{1}{4} \log \frac{m_A^2}{m_\tau^2} \right)\,,\nonumber\\\
c_{L}^{1-{\rm loop},H^+} &= -\frac{ m_\tau \eta_{\tau\mu}}{\sqrt2 vm_{H^+}^2}\frac{\hat\eta_\tau}{12} \,, \\
c_{R}^{1-{\rm loop},H^+} &= \frac{m_\mu}{m_\tau} c_{L}^{1-{\rm loop},H^+} \,.
\end{align}
Note that, since $\hat\eta_\mu$ might be much larger than $\hat\eta_\tau$, one should consider also diagrams with internal muons in the loops. Since in our numerical analysis we take both $\hat\eta_\tau=0$ and $\hat\eta_\mu=0$, we do not write these corrections explicitly.

The dominant two-loop effects are of the Barr-Zee type involving the top quark loop. In addition, the CP even scalars also contribute Barr-Zee type diagrams with the $W$ boson loop. Contributions involving internal $Z$-boson exchange are suppressed compared to those involving photons and we neglect them. The relevant  contributions to the decay amplitude can thus be written as~\cite{Chang:1993kw, Davidson:2010xv, Goudelis:2011un, Harnik:2012pb}
\begin{align}
c_{L,R}^{2-{\rm loop, t},h} & = \frac{\alpha_{\rm EM}}{3\sqrt2\pi} \frac{\mathcal S'(z_{t/h})}{v m_\tau}\left(c_{\ahm}+\hat\eta_t s_{\ahm}\right) \eta_{\tau\mu,\mu\tau}s_{\ahm}\,,\\
c_{L,R}^{2-{\rm loop, t},H} & = \frac{\alpha_{\rm EM}}{3\sqrt2\pi} \frac{\mathcal S'(z_{t/H})}{v m_\tau}\left(-s_{\ahm}+\hat\eta_t c_{\ahm}\right) \eta_{\tau\mu,\mu\tau}c_{\ahm}\,,\\
c_{L,R}^{2-{\rm loop, t},A} & = \frac{\alpha_{\rm EM}}{3\sqrt2\pi} \frac{\mathcal P'(z_{t/A})}{v m_\tau} \hat\eta_t \eta_{\tau\mu,\mu\tau}\,, \\
c_{L,R}^{2-{\rm loop, W},h} & = +\frac{4\sqrt2\alpha_{\rm EM}}{\pi} \frac{\mathcal A'(z_{W/h})}{v \,m_\tau} c_{\ahm} s_{\ahm} \eta_{\tau\mu,\mu\tau} \,,\\
c_{L,R}^{2-{\rm loop, W},H} & = -\frac{4\sqrt2\alpha_{\rm EM}}{\pi} \frac{\mathcal A'(z_{W/H})}{v \,m_\tau} c_{\ahm} s_{\ahm}  \eta_{\tau\mu,\mu\tau} \,,
\end{align}
where the relevant two-loop functions are given by
\begin{align}
\mathcal S'(z) &=  \mathcal P' (z) (1-2z) + z \log z + 2 z \,, \\
\mathcal A'(z) &= \frac{1}{4(1-4z)} \left[   \mathcal P'(z) (96 z^2 -158z+35)  \right.\nonumber\\
&\left.+  3(5-16z) z \log z + 24 z (1-4z) \right]\,,\\
  \mathcal P'(z) & = \frac{z}{\sqrt{1-4 z}}  \left[\text{Li}_2\left(\frac{2}{1-\sqrt{1-4
   z}}\right) \right. \nonumber\\
& \left.   - \text{Li}_2\left(\frac{2}{1+\sqrt{1-4
   z}}\right) \right]
   +\frac{2 \, z \, \log
  z \cot ^{-1}\left(\sqrt{4 z-1}\right)}{\sqrt{4
   z-1}}\,.
\end{align}
A full calculation of the two-loop charged Higgs contributions has not yet appeared in the literature and is beyond the scope of this work.

Additional contributions to the dipole operators arise from the full UV model which generates $c^{H,A}_{\gamma\gamma}$.
These are generated at the scale $m_S$ by mixing with $Q_{ijBB}$, and, at lower scales, by integrating out the heavy 2HDM scalars.
We use the results of~\cite{Alonso:2013hga}, to find
\begin{align}
\dot c_R&=\gamma\frac{\eta^{\mu\tau*}}{y_\tau}\left(c_{12BB}+ic_{12\tilde{B}B}\right)
\,,\nonumber\\
\dot c_L&=\gamma\frac{\eta_{\tau\mu}}{y_\tau}\left(c_{12BB}-ic_{12\tilde{B}B}\right)\,,
\end{align}
with $\gamma=6\sqrt{2}\pi^2$, and
\begin{align}
c_{12\tilde{B}B}=\frac{3}{2}c_{12BB}=\frac{\alpha_{\rm EM}}{c_W^2\pi v^2}c_{A\gamma}\,.
\end{align}
This mixing then induces
\begin{align}
c_{L}=\frac{\gamma\alpha_{\rm EM}c_{A\gamma}}{\pi c_W^2v^2}\frac{\eta_{\tau\mu}}{y_\tau}&\left[\log\left(\frac{\Lambda}{m_S}\right)\left(\frac{2}{3}-i\right)\right.\nonumber\\
&\left.+\frac{2}{3}\sin\ahm\log\left(\frac{m_S}{m_h}\right)\right]\,,\nonumber\\
c_{R}=\frac{\gamma\alpha_{\rm EM}c_{A\gamma}}{\pi c_W^2v^2}\frac{\eta_{\mu\tau}^{*}}{y_\tau}&\left[\log\left(\frac{\Lambda}{m_S}\right)\left(\frac{2}{3}+i\right)\right.\nonumber\\
&\left.+\frac{2}{3}\sin\ahm\log\left(\frac{m_S}{m_h}\right)\right]\,.
\end{align}
Here $\Lambda$  is the mass scale of heavy dynamics generating the EFT operators. We neglect the weak running of $c_\gamma^A$ and further finite corrections
which are not logarithmically enhanced.

\subsection{Additional leptonic constraints}

A comment is in order regarding other lepton flavor violating processes. The $\bar\psi\gamma^\mu\psi H^\dagger D_\mu H$ operators, which lead to lepton flavor changing $Z$ couplings, do not mix with the $Q_{ijBB}$ operators at one-loop. Hence, assuming it is not generated by the heavy dynamics, there are no significant contributions to, {\it e.g.}, $\tau\rightarrow3\mu$ process. In the absence of $\eta_{\mu\mu}$, the relevant tree-level amplitudes are suppressed by the flavor-diagonal muon Yukawa and the small mixing angle. At the one-loop level, it was shown, {\it e.g.}, in~\cite{Harnik:2012pb}, that the bounds on the dipole operator arising from the $\tau\to\mu\gamma$ constraints are stronger than the ones arising from this process or from the muon dipole moments.

If both $\eta_{\tau\mu}$ and $\eta_{\mu\tau}$ exist, the $\tau\rightarrow\mu\nu\bar\nu$ decay deviates from its SM prediction~\cite{Omura:2015xcg},
\begin{align}
\Gamma\left(\tau\rightarrow\mu\nu\bar\nu\right)&=\Gamma\left(\tau\rightarrow\mu\nu\bar\nu\right)^{\rm SM}\left(1+\frac{|\eta_{\mu\tau}|^2|\eta_{\tau\mu}|^2}{32G_F^2m_{H^+}^4}\right)
\end{align}
Current experimental constraints on the lepton universality decays read~\cite{Aubert:2009qj}
\begin{align}
|\eta_{\mu\tau}\eta_{\tau\mu}|\lesssim3.2\,,
\end{align}
at $95\%$ C.L.. The presence of these two couplings might also generate an electric dipole moment for the muon. Current bounds read, at $95\%$ C.L.~\cite{Agashe:2014kda},
\beqa
-0.36\lesssim{\rm Im}\left[\eta_{\mu\tau}\eta_{\tau\mu}\right]\sin^2\ahm\lesssim0.40\,.
\eeqa
The muon magnetic moment measurement, on the other hand, exhibits a $3\,\sigma$ discrepancy with respect to its SM prediction. Although we do not aim to explain this discrepancy, it might be accommodated within the scenario we consider, provided that (c.f.~\cite{Omura:2015nja})
\beqa
2.6\times10^{-3}\lesssim{\rm Re}\left[\eta_{\mu\tau}\eta_{\tau\mu}\right]\sin^2\ahm\lesssim8.8\times10^{-3}\,.
\eeqa
Referring to the latter as an upper bound, the muon magnetic and electric moments can be combined to give
\begin{align}
|\eta_{\mu\tau}\eta_{\tau\mu}|\sin^2\ahm\lesssim0.4\,,
\end{align}

\subsection{Higgs data}

The measurements of the light $h$ in the various decay modes are also considered.
The $h\to\gamma\gamma$ decay rate depends, however, on additional EFT parameter, $c_{11BB}$, which is unconstrained by the di-photon signal at $750$~GeV. We therefore analyze the Higgs data in two ways. First, we leave out the di-photon measurement at 125~GeV, allowing for cancelation between the EFT contributions and the SM ones. Alternatively, one can assume that the UV contributions to $h\to\gamma\gamma$ are negligible, by considering only the top and $W$ boson loops in the $h\gamma\gamma$ coupling. These two approaches are somehow orthogonal, and capture different types of UV completed theories. Note that simultaneous cancelation of the NP contributions both in the $h\to\gamma\gamma$ process, and the $\tau\to\mu\gamma$ process, is impossible.

\section{Results}
\label{sec:results}

As concerns $h$, $\sin\ahm$ affects all production and decay rates, $\hat\eta_{t}$ affects the decay rate of $h\to gg$ and the ggF production rate, while $\hat\eta_b$ affects mainly the total width of $h$. We note that in the allowed parameter space $|\hat\eta_b \sin \alpha_{vh}| \lesssim 0.4$ is required by the Higgs measurements, rendering the bottom loop contribution to the ggF negligible.

As concerns $S$, $\sin\ahm$ affects all production and decay rates. $\hat\eta_t$ affects the $S\to gg$ decay rate and the ggF production rate of $S$. It further affects the decay rate of of $S\to t\bar t$ and its total decay width. $\hat\eta_b$ affects the $S\to b\bar b$ decay rate and the  $b\bar b S$ production (with milder effect on its total width, in the relevant region of $\hat\eta_b$).

Both an upper and a lower bound on $\sin\ahm$ are implied from requiring $\Gamma_{H,A}<45$~GeV. At large $\sin\ahm$, the total width is dominated by the $A\to hZ$ decay. At small $\sin\ahm$, the total width is dominated by the $S\to\tau\mu$ decay. Combining the two, we find,
\begin{align}
2\times10^{-3}\lesssim\sin\ahm\lesssim0.22\,,
\end{align}
independently of $\hat\eta_{t,b}$, which also satisfy EWPM at the 95\% C.L..

\begin{figure}[h!]
  \begin{center}
      \subfigure[]{\scalebox{0.995}{\includegraphics[width=2.5in]{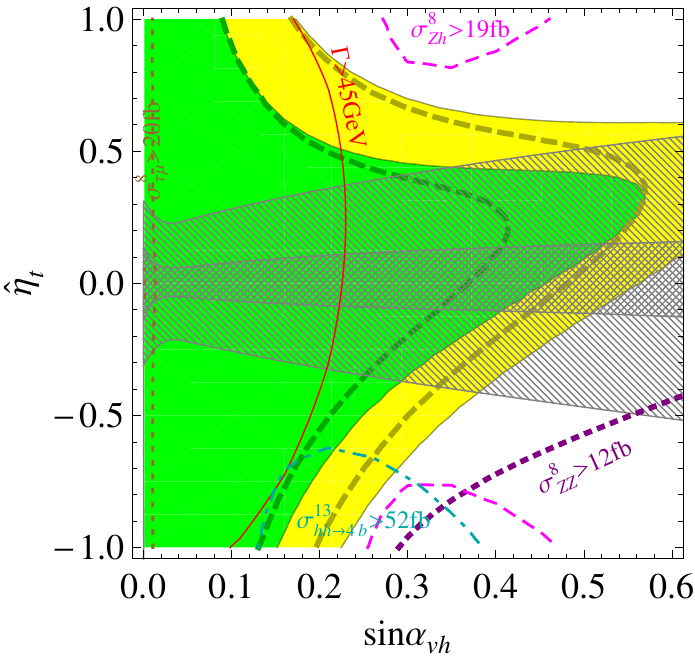}}}
      \subfigure[]{\scalebox{0.995}{\includegraphics[width=2.5in]{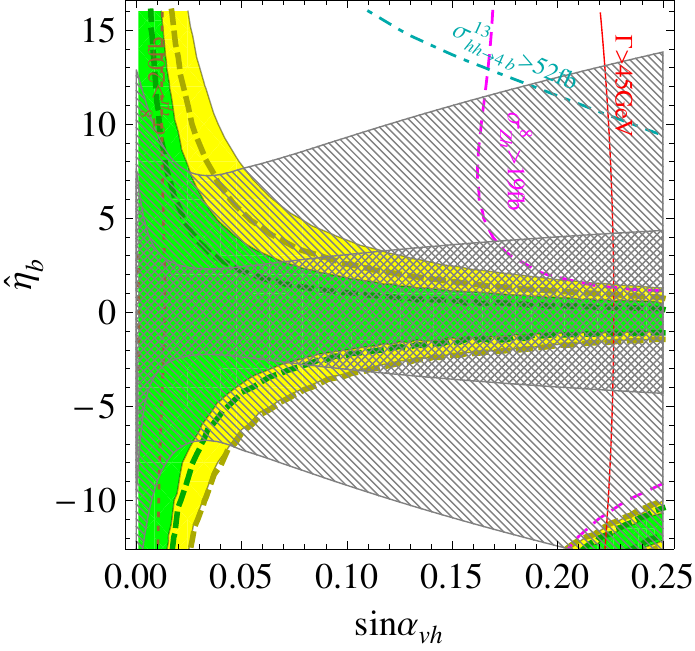}}}
      \caption{\label{fig:results_1} The 2HDM allowed 95\% C.L. region in the (a) $\sin\ahm-\hat\eta_t$ plane (with $\hat\eta_b=0$) and (b) $\sin\ahm-\hat\eta_b$ plane (with $\hat\eta_t=0$). The LHC Higgs data (excluding the $h\to\gamma\gamma$ measurements) allows the green (yellow) region within 68\% (95\%)~C.L.. The corresponding dashed contours include the $h\to\gamma\gamma$ measurement. In the inner (outer) grey region $r^{13/8}_{\gamma\gamma}<3\,(2)$, which is disfavored by the di-photon search at 8~TeV. Other constraints are explained in the figure.}
  \end{center}
\end{figure}

\begin{figure}[h!]
  \begin{center}
      \subfigure[]{\scalebox{0.99}{\includegraphics[width=2.5in]{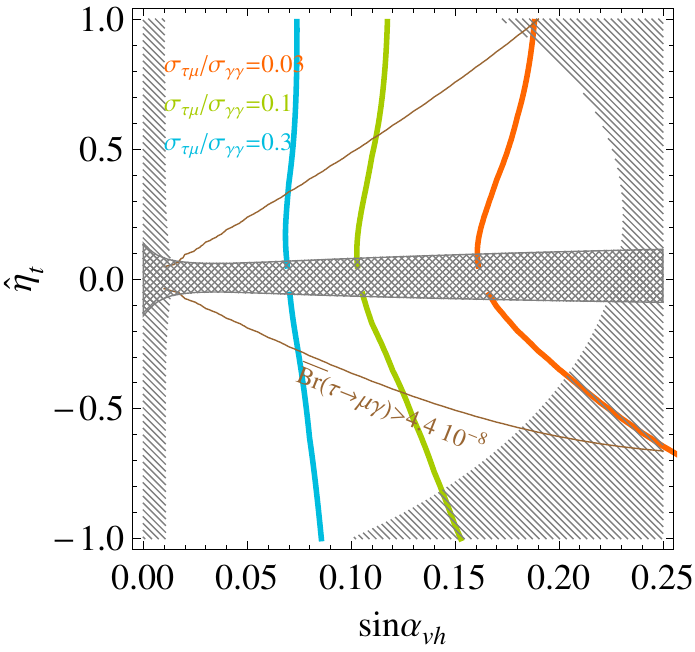}}}
      \subfigure[]{\scalebox{0.99}{\includegraphics[width=2.5in]{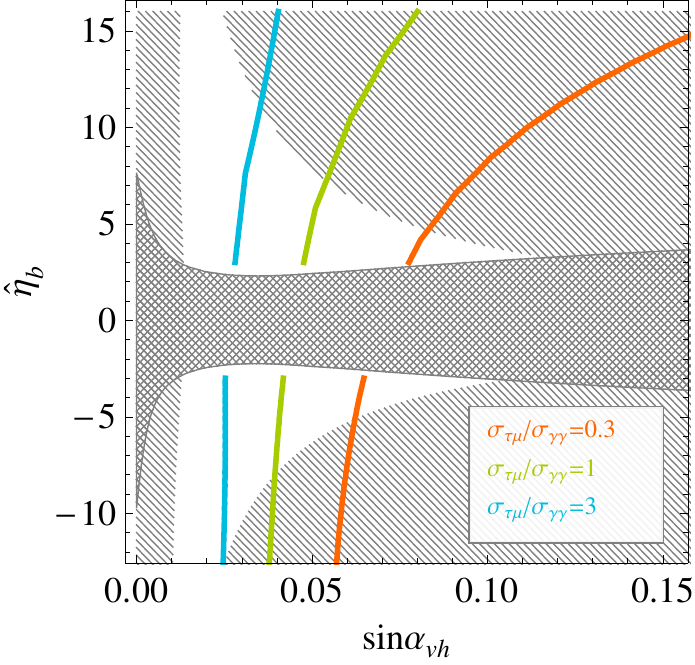}}}
      \caption{\label{fig:results_2} The expected $S\to\tau\mu$ signal of the 2HDM in the (a) $\sin\ahm-\hat\eta_t$ plane (with $\hat\eta_b=0$) and (b) $\sin\ahm-\hat\eta_b$ plane (with $\hat\eta_t=0$). Excluded region is shaded, see the text for more details.}
        \end{center}
\end{figure}

Our numerical results are presented in Figs.~\ref{fig:results_1},\ref{fig:results_2} for the $\hat\eta_t-\sin\ahm$ and $\hat\eta_b-\sin\ahm$ parameter spaces.

Fig.~\ref{fig:results_1} shows the constraints from the various experimental results described above.
The LHC Higgs data allows the green (yellow) region within 68\% (95\%)~C.L.. For this, we leave out the $h\to\gamma\gamma$ measurements which is affected by additional (unconstrained) EFT parameter, $c_{11BB}$ (see discussion in the previous section). We further include this measurement, assuming $c_{11BB}=0$, in the dashed lines (using the same color scheme). 

The grey dashed region shows the parameter region disfavored by the 8~TeV di-photon search. Specifically, in the
inner (outer) dashed region we find $r^{13/8}_{\gamma\gamma}<3\,(2)$. Other constraints are explained in the figure.

Fig.~\ref{fig:results_2} shows, in the allowed (white) region, our predictions for the $S\to\tau\mu$ signals. To be conservative, we exclude in this figure only $r^{13/8}_{\gamma\gamma}<3$. The $\tau\to\mu\gamma$ line shown in this plot includes only the known 2HDM contributions to this process, and should be taken only as an order of magnitude estimate. 
As for the $pp\to S\to\tau\mu$ process, we find that a signal of $\mathcal{O}$(1-10)~fb is possible within our framework. 

\section{The phenomenology of $H^+$}
\label{sec:charged}
In absence of $\eta^{u,d}_{ij}$ couplings to light quarks, the dominant $H^\pm$ production mechanism at the LHC is the associated production with a top quark:
\begin{align}
\sigma(pp \to H^- t) & = \sigma(pp \to H^+ \bar t) = |\hat\eta_t|^2 ~63.0~(7.96) {~\rm fb} \,,
\end{align}
at 13 (8) TeV. These are evaluated with MG5~\cite{Alwall:2014hca} using NLO NNPDF2.3~\cite{Ball:2012cx} pdf set.
The corresponding values for $\hat\eta_b$ can be deduced by replacing $\hat\eta_t\to\hat\eta_bm_b/m_t$.
The dependence on $\hat\eta_{t,b}$ relates between the production cross sections of the charged Higgs and the neutral scalars. These further generate, at the one-loop order, a non-zero $\eta_{bc}$. This contribution is further suppressed by weak coupling and the small CKM elements $|V_{cb}|,|V_{ts}|$, as well as the bottom quark mass, and therefore can be safely neglected.

As for the $H^\pm$ decays,
\begin{enumerate}
\item In the absence of $\eta_{\ell\ell}$, the leptonic decay modes of the charged Higgs are
\begin{align}
\Gamma(H^+ \to \tau^+ \nu) & = \frac{|\eta_{\tau\mu}|^2}{16\pi} m_{H^+}\,, \\
\Gamma(H^+ \to \mu^+ \nu) & = \frac{|\eta_{\mu\tau}|^2}{16\pi} m_{H^+}\,.
\end{align}
These decays provide a window into the chirality structure of the LFV coupling.
\item The decay to quarks is dominated by
\begin{align}
\Gamma(H^+ \to t \bar b) & = \frac{3|\hat\eta_{t,b}|^2}{16\pi}\left(\frac{2m_t^2}{v^2}\right)^2 m_{H^+} \beta[2m_{t,b},m_{H^+}]\,.
\end{align}
\item The decay into final bosons reads
\begin{align}
\Gamma(H^+ \to h W^+) & = \frac{G_F m_{H^+}^3 \sin^2\ahm}{8\sqrt{2}\pi}\lambda\left[m_h^2,m_W^2,m_{H^+}^2\right]^{3/2}\,.
\end{align}
\end{enumerate}

Let us define, similarly to our definition in the neutral scalar sector,
\begin{align}
R^{H^+/A}_{XYZ}&\equiv \frac{\Gamma(H^+\to XY)}{\Gamma(A\to XZ)}\,.
\end{align}
Clearly, SU(2) invariance relates between the decays of the charged Higgs to those of the pseudoscalar, such that
\begin{align}
R^{H^+/A}_{hW^+Z}\simeq R^{H^+/A}_{t\bar b\bar t} = R^{H^+/A}_{\tau^+\nu\mu^-}+R^{H^+/A}_{\mu^+\nu\tau^-} =1
\end{align}

The $hW^\pm$ signature is complex, and we are not aware of an experimental analysis searching for a $tWh$ final state. The decay into $tb$ was searched for $m_{H^+}<600$~GeV~\cite{Aad:2015typ}, reaching a sensitivity of 200~fb for this mass. An improvement of an order of magnitude would be needed to probe some of the parameter space in this channel at $750$~GeV.

\begin{figure}[h!]
	\begin{center}
	\includegraphics[height=2in]{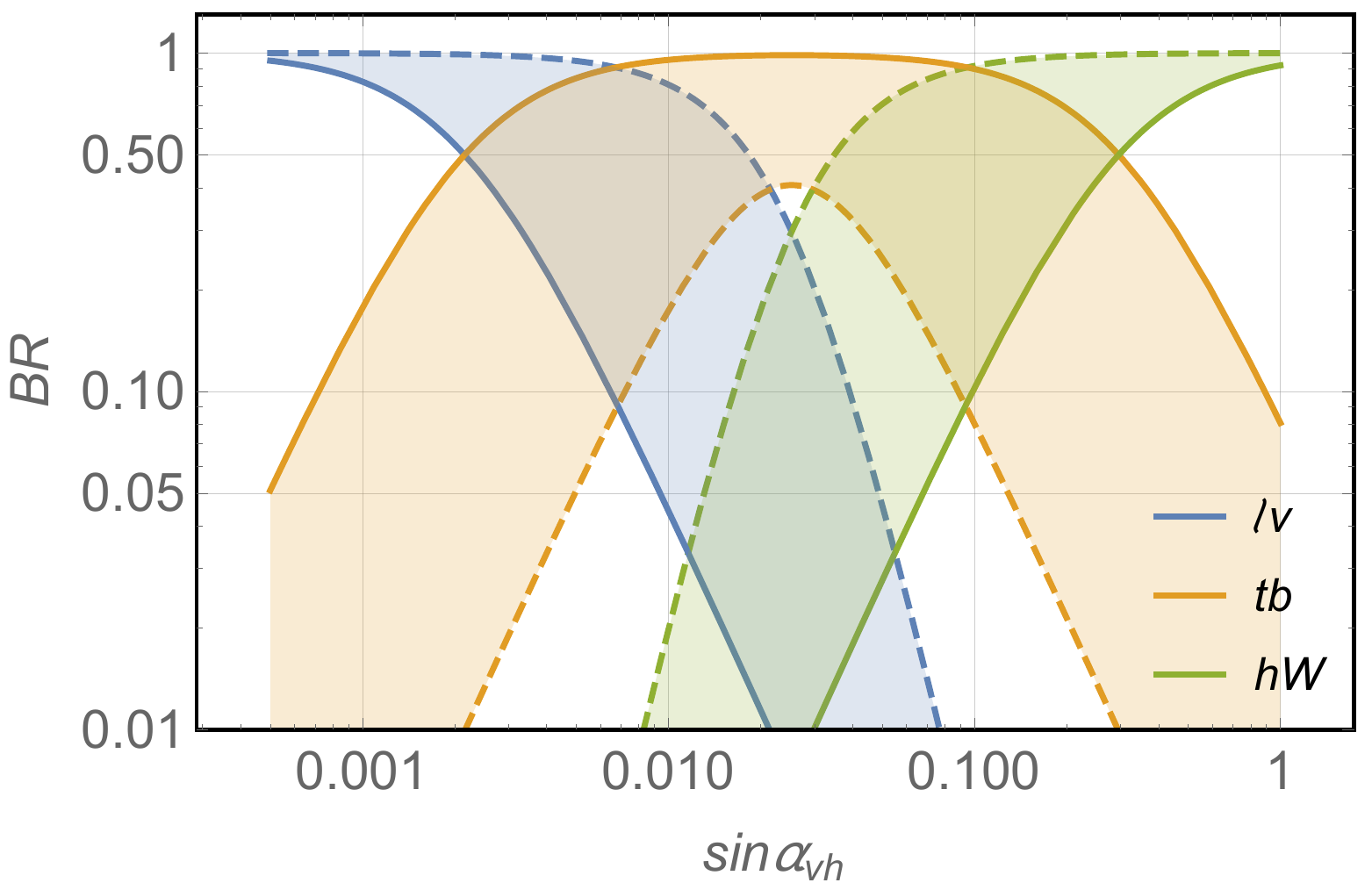}
	\caption{\label{fig:2} $H^\pm$ branching ratios as a function of $\sin\ahm$, with $\hat\eta_t=1$ (solid lines) and $\hat\eta_t=0.1$ (dashed lines). The leptonic channels are sumed over $\tau^\pm\nu$ and $\mu^\pm\nu$. The corresponding values for $\hat\eta_b$ can be deduced by replacing $\hat\eta_t\to\hat\eta_bm_b/m_t$.}
	\end{center}
	\end{figure}

\begin{figure}[h!]
	\begin{center}
	\includegraphics[height=2in]{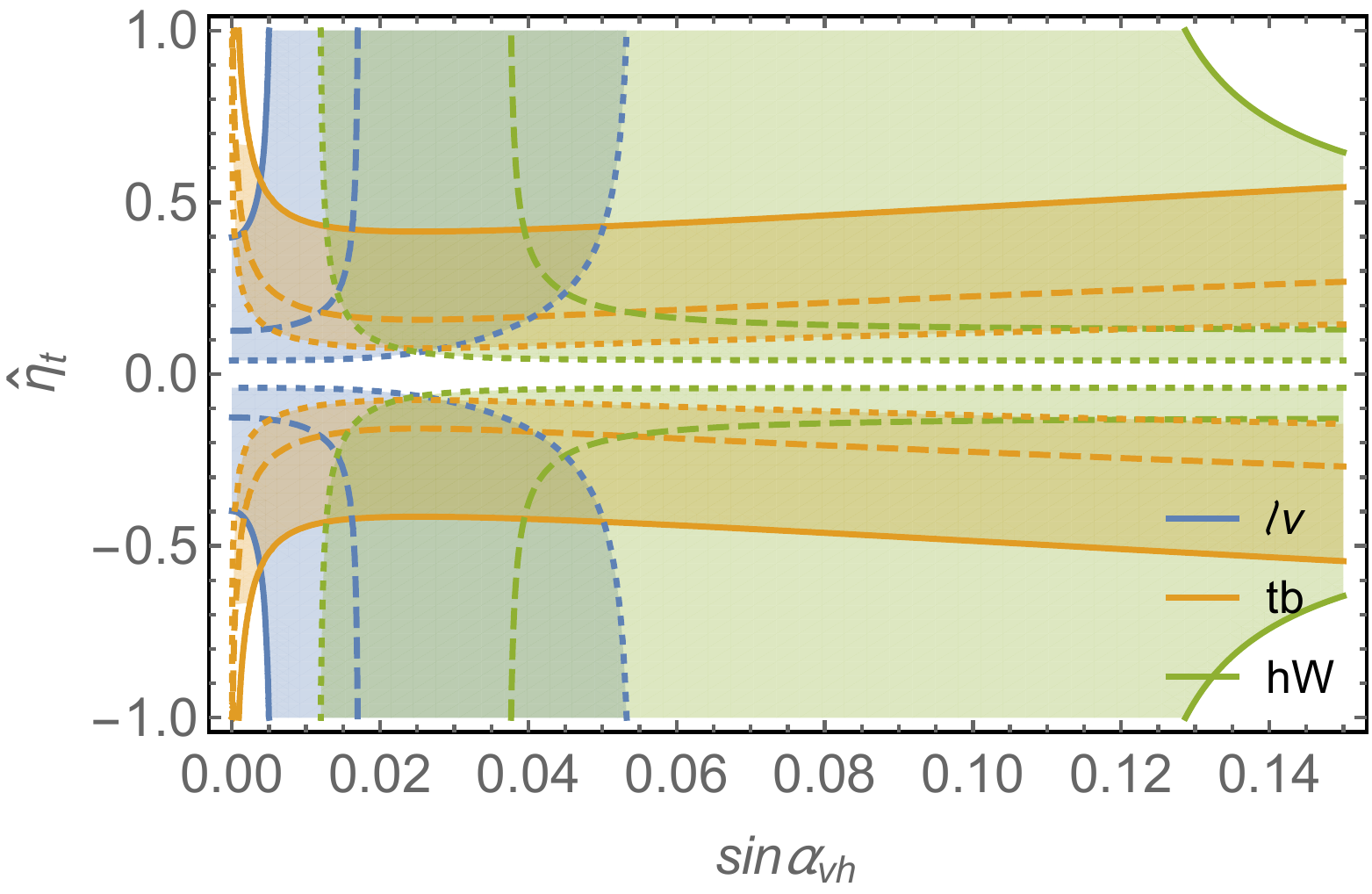}
	\caption{\label{fig:3} The expected signal at 13~TeV of the associated $H^+\bar t$ production in the various decay modes. Contours are shown for 10, 1 and 0.1 fb signals in solid, dashed and dotted lines, respectively. The leptonic channels are sumed over $\tau^+\nu$ and $\mu^+\nu$. The corresponding values for $\hat\eta_b$ can be deduced by replacing $\hat\eta_t\to\hat\eta_bm_b/m_t$.}
	\end{center}
	\end{figure}

As concerns the leptonic modes, the current bound reads, at $95\%$ C.L.~\cite{Aad:2014kga,Aaboud:2016dig},
\begin{align}
\sum_{+-}\sigma_{13}(pp\to tH^\pm)\times{\rm BR}(H^\pm\rightarrow\tau^\pm\nu)&\lesssim25{\rm ~fb}\,,
\end{align}
which implies
\begin{align}
(2m_{t,b}^2/v^2)|\hat\eta_{t,b}|^2{\rm BR}(H^+\rightarrow\tau^+\nu)\lesssim0.05\,.
\end{align}
The various branching ratios of the charged Higgs are shown in Fig.~\ref{fig:2}.
Fig.~\ref{fig:3} shows the expected signals for the various decay modes of the charged Higgs. The contours for
10, 1 and 0.1 fb signals are plotted in solid, dashed and dotted lines, respectively. The $\ell\nu$ mode is summed over the $\tau\nu$ and $\mu\nu$ decays. We stress that the $\mu\nu$ final state would be a clean signature of the $H^\pm$ decay in the presence of $\eta_{\mu\tau}$.

%
%

\section{Conclusions}
\label{sec:conclusions}
If the di-photon resonance at 750 GeV is experimentally established, a possible interpretation would be that it constitutes of the neutral members of a second Higgs doublet. In such a case, it is plausible that the Yukawa couplings of the light Higgs at 125 GeV are not purely diagonal and, in particular, the $h\to\tau\mu$ decay at observable rate is allowed. We analyzed the lessons from present data that follow from the 2HDM interpretation of the $S$ resonance and from assuming that ${\rm BR}(h\to\tau\mu)\sim0.01$.

Our main conclusions are the following:
\begin{itemize}
\item The rate of $S\to\tau\mu$ could be comparable to the rate of $S\to\gamma\gamma$. In fact, in regions of the parameter space where $h$ is very close to the direction of the VEV ($\sin\ahm\sim0.002$), $S\to\tau\mu$ can be the dominant decay mode.
\item $S$ should also be searched for in various di-boson final states: $VV$, $Zh$ and $hh$. In regions where $h$ is not very close to the direction of the VEV  ($\sin\ahm\sim0.2$), $S\to Zh$ can be the dominant decay mode.
\item The charged Higgs $H^\pm$ should be not far in mass from 750 GeV. It should be searched for, in addition to the $tb$ and $\tau\nu$ modes, in the $Wh$ and $\mu\nu$ modes. The balance between the $\tau\nu$ and $\mu\nu$ branching ratios can provide a unique window into the chirality structure of lepton flavor violating decays.
\end{itemize}

\begin{acknowledgments}
JFK and YN thank Ezequiel Alvarez, the organizer of the ``Voyages Beyond the SM" workshop, and the participants of the workshop, for stimulating atmosphere and discussions during the workshop. We thank Liron Barak for useful discussions.
JFK acknowledges the financial support from the Slovenian Research Agency (research core funding No. P1-0035).
YN is the Amos de-Shalit chair of theoretical physics. YN is supported by the I-CORE program of the Planning and Budgeting Committee and the Israel Science Foundation (grant number 1937/12), and by a grant from the United States-Israel Binational Science Foundation (BSF), Jerusalem, Israel.
\end{acknowledgments}
%
%
%
%

\end{document}